\documentclass[prd,twocolumn,a4paper,superscriptaddress,floatfix]{revtex4}
\usepackage{graphicx}
\begin{document}

\newcommand{\be}{\begin{equation}}
\newcommand{\ee}{\end{equation}}
\newcommand{\bq}{\begin{eqnarray}}
\newcommand{\eq}{\end{eqnarray}}
\newcommand{\bsq}{\begin{subequations}}
\newcommand{\esq}{\end{subequations}}
\newcommand{\bc}{\begin{center}}
\newcommand{\ec}{\end{center}}
\newcommand {\R}{{\mathcal R}}
\newcommand{\al}{\alpha}
\newcommand\lsim{\mathrel{\rlap{\lower4pt\hbox{\hskip1pt$\sim$}} \raise1pt\hbox{$<$}}}
\newcommand\gsim{\mathrel{\rlap{\lower4pt\hbox{\hskip1pt$\sim$}} \raise1pt\hbox{$>$}}}

\title{ Dynamical and Thermodynamical Aspects \\ of Interacting Kerr Black Holes}

\author{Miguel S. Costa}
\email[Electronic address: ]{miguelc@fc.up.pt}
\author{Carlos A.R. Herdeiro}
\email[Electronic address: ]{crherdei@fc.up.pt}
\author{Carmen Rebelo}
\email[Electronic address: ]{mrebelo@fc.up.pt}
\affiliation{Centro de F\'{\i}sica do Porto e Departamento de F\'{\i}sica da Faculdade de Ci\^encias da Universidade do Porto, Rua do Campo Alegre 687, 4169-007 Porto, Portugal}

\date{March 2009}

\begin{abstract}
We consider the asymptotically flat double-Kerr solution for two equal mass black holes with either the same or  opposite angular momentum and with a massless strut between them. For fixed angular momentum and mass, the angular velocity of two co-rotating Kerr black holes decreases as they approach one another, from the Kerr value at infinite separation to the value of a single Kerr black hole with twice the mass and the angular momentum at the horizons merging limit. We show that the ratio $J/M^2$ for extremal co-rotating Kerr black holes varies from unity at infinite separation to two at the merging limit. These results are interpreted in terms of rotational dragging and compared with the case of counter-rotating Kerr black holes. We then analyse the merging of ergo-regions. In the co-rotating case the merger point occurs at an angle of $\pi/2$, in agreement with recent general arguments. In the counter-rotating case the ergo-regions never merge. We study the horizon geometry for both cases as a function of the distance and provide embedding diagrams. Finally, we study the thermodynamical evolution of the co-rotating double Kerr system, showing that, in the canonical ensemble, it is thermodynamically stable for fast spinning black holes. As for single Kerr black holes the stable and unstable phases are separated by a second order phase transition. We show that for large fixed angular momentum two Kerr black holes reach a minimum distance,
before horizon merging has occurred, where the thermodynamical approximation breaks down. We also consider the micro-canonical ensemble to study the maximal energy that can be extracted from the 
double Kerr system as a function of the separation between the black holes.
\end{abstract}
\maketitle

\section{\label{sint}Introduction}
Understanding the interaction between two Kerr black holes is of crucial importance in connecting the General Theory of Relativity with observations of astrophysical black hole binaries . One expects that the full dynamical evolution of a binary system of Kerr black holes, in particular the merger stage, can only be tackled in numerical relativity. Despite the remarkable recent progress, which has allowed one to infer the properties of the final state in binary black hole coalescence (see \cite{Rezzolla} for a review), there are still issues which the current numerical codes have not yet addressed. For instance, the codes do not keep track of the horizon angular velocities of individual black holes, but only of their angular momentum \cite{Dreyer:04}.  The angular velocities are of physical interest because they are associated to the existence of an ergo-region which, in turn, is associated to potentially observable phenomena such as superradiance.

As a first step for understanding this and other phenomena in a realistic binary system, one can consider a quasi-static approximation, where relative velocity effects are neglected and the black holes are kept in equilibrium by an external force. Studies of such systems include the double-Schwarzschild solution \cite{Weyl, IK, CP} and the \textit{double Kerr solution} \cite{KN}. The latter is an exact solution of the four dimensional vacuum Einstein equations which, for some region in parameter space, describes two Kerr black holes rotating in the same plane, in an asymptotically flat spacetime and with a massless strut between them, which provides the equilibrium.  For instance, we have recently shown \cite{HR} that for a special case of the double Kerr system (for equal Komar masses $M_1=M=M_2$ and opposite Komar angular momenta $J_1=J=-J_2$) one can provide an exact formula in terms of the physical parameters for the angular velocity of the black holes:
\begin{equation} 
\Omega_1=\frac{J}{2M\left(M^2+\sqrt{M^4-J^2\frac{\zeta -2M}{\zeta+2M}}\right)}\frac{\zeta -2M}{\zeta+2M}=-\Omega_2 \ , 
\label{counter1} 
\end{equation}
where $\zeta$ is the coordinate distance between the black holes in Weyl canonical coordinates. This angular velocity varies from the Kerr value as $\zeta \rightarrow +\infty$, to zero as $\zeta\rightarrow 2M$, corresponding to the touching of the two horizons. Such \textit{slow down} of the angular velocity is very intuitively interpreted as a consequence of the rotational dragging that the black holes exert on each other.

As a consequence of (\ref{counter1}), one may suspect that, in this \textit{counter-rotating} system, the ergo-regions never merge because, as one decreases the distance between the black holes, they 
will become closer and closer to the horizons, collapsing onto the horizons when the horizons themselves touch. This shall be explicitly shown in this paper. One should note that in this system the Kerr upper bound for the angular momentum depends on $\zeta$ \cite{HR}:
\begin{equation} 
\frac{|J|}{M^2}=\sqrt{\frac{\zeta+2M}{\zeta-2M}} \ , \label{counter2} 
\end{equation}
varying from the Kerr value as $\zeta \rightarrow +\infty$, to infinity as $\zeta\rightarrow 2M$. Thus, the ergo-regions never touch even tough the individual black holes can have an \textit{arbitrarily large} angular momentum, for fixed mass, if they are sufficiently close. The ``trick'' is that, despite the arbitrarily large angular momentum, the angular velocity is always sufficiently small as to allow a black hole horizon. This example, therefore, makes crystal clear that it is the angular velocity, and not the angular momentum, that determines the existence of a horizon, for an uncharged, rotating black hole, an intuitive observation if one invokes the membrane paradigm \cite{Membrane}.

If one considers the \textit{co-rotating} system (equal Komar masses $M_1=M=M_2$ and angular momenta $J_1=J=J_2$), one  might naively expect an opposite behaviour. That is, as they approach one another, the mutual rotational dragging should \textit{speed up} the horizon angular velocity of either black hole. Reasoning along the same lines as in the last paragraph, one could then expect that the extremality limit would \textit{decrease} as compared to the usual Kerr value. These expectations, however, immediately lead us into trouble. A first argument is the following. If this were true, two under-extreme Kerr black holes (no strut between them), dropped from a certain distance with zero velocity could become extreme and over-extreme as they would fall into one another. Indeed, the initial distance and angular momenta could be tuned such that velocity effects are as small as one wishes, so that one could trust the results of the double-Kerr solution as a quasi-static approximation. This would clearly violate cosmic censorship and, potentially, the second law of thermodynamics. A second argument is that, in this co-rotating case, one expects a single Kerr black hole with twice the mass and twice the angular momentum to arise when the black holes merge. This will indeed be true, the reason being clearly understood from the rod structure of the co-rotating double Kerr solution, to be described below. For a single Kerr black hole, the angular velocity as a function of mass and angular momentum obeys
\[
\Omega(2M,2J)=\frac{\Omega(M,J/2)}{2}<\Omega(M,J) \ . 
\]
Thus, if the double Kerr co-rotating solution interpolates between two Kerr black holes of mass $M$ and spin $J$ and one Kerr black hole of mass $2M$ and spin $2J$, the angular velocity of the horizon must \textit{slow down} as we decrease the distance between them. Consequently, the extremality limit should \textit{increase} as compared to the usual Kerr value. It is simple to see that a co-rotating double Kerr system that interpolates (fixing $M,J$) between a final extremal black hole and two initial Kerr objects, requires the latter to obey $J=2M^2$. Thus, we expect the extremality limit to vary between the Kerr value, asymptotically, and $J=2M^2$ just before merging. 

The analysis of the co-rotating double-Kerr system is more  involved and no simple expressions analogue to (\ref{counter1}) and (\ref{counter2}) could be obtained. Nevertheless, we will exhibit all physical quantities in a way analogous to the ones of the counter-rotating case. We shall show numerically that, for fixed mass and angular momentum, the angular velocity of the two black holes indeed slows down as they approach; and that the ratio $J/M^2$ for extremal black holes grows from unity, asymptotically, to two, in the merging limit. It follows that, in this co-rotating case, we expect the ergo-regions of the two black holes to indeed merge, and eventually become a single one, as they approach.

\begin{figure}
\includegraphics[totalheight=3.5in,viewport=25 78 650 717,clip]{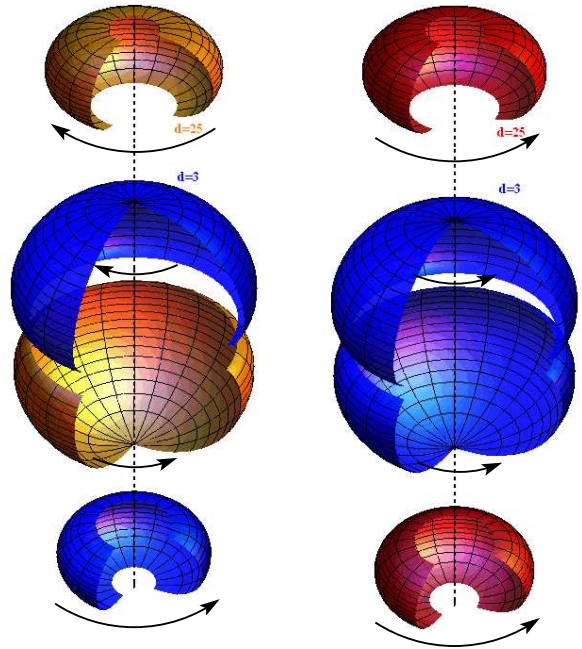}
\begin{picture}(0,0)(0,0)
\end{picture}
\caption{\label{embed} Horizon embedding for the counter-rotating (left) and co-rotating (right) double Kerr system, for two different  values of the physical distance $d$  and for $M=1=J$.}
\end{figure}

\begin{figure}
\includegraphics[width=3.2in]{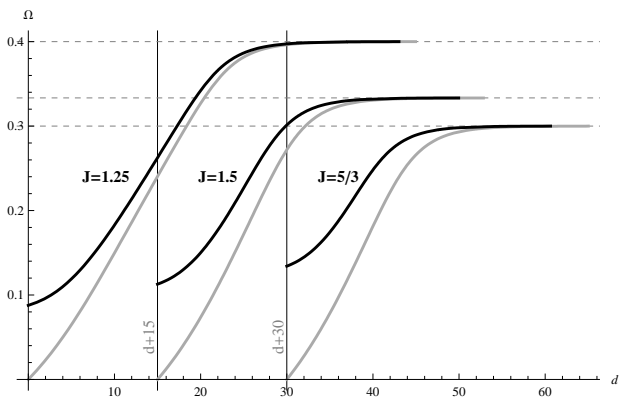}
\begin{picture}(0,0)(0,0)
\end{picture}
\caption{\label{angcomparison} Horizon angular velocity in terms of the physical distance for $M=1$ and $J=1.25,1.5,5/3$, for both the counter-rotating (grey) and co-rotating (black) cases. We have chosen values of $J/M^2$ greater than unity to enhance the difference between the two cases. For these choices the solutions describe black holes only up to some maximal physical distance.}
\end{figure}

The behaviour obtained in the co-rotating system can still be interpreted in terms of rotational dragging. The point is that each black hole is simultaneously \textit{dragging} and \textit{being dragged}. To drag some material object effectively increases the moment of inertia of the black hole and hence decreases its angular velocity. In the counter-rotating Kerr system, the dragging exerted \textit{by} and \textit{upon} each black hole work with the same purpose: slowing down each one of them. But in the co-rotating system they work with opposite goals: on the one hand, \textit{dragging} the other black hole slows down the black hole which is dragging; on the other hand, \textit{being dragged} by the other black hole speeds up the black hole which is being dragged. The  observed behaviour shows that it is the first effect that is the dominant one, and this was exactly the effect that was being neglect in the naive expectations mentioned above.  This interpretation is qualitatively confirmed by the fact that, for the same $M,J$ and physical distance $d$, the angular velocity is smaller in the counter-rotating, where both effects slow down each horizon, than in the co-rotating case - Fig. \ref{angcomparison}.

Since in the co-rotating case we expect the two ergo-spheres to merge we may ask what is the merger angle at which this happens. Recently, a universal behaviour for mergers of ergoregions has been observed \cite{ELVANG}. For $D+1$ dimensional vacuum solutions of the Weyl class with rotation in a single plane, the merger angle is always given by
\[ \theta_m=2{\rm arccot}\sqrt{\delta -1} \ , \]
where $\delta=D-p$ and the merger is extended along $p$ spatial dimensions. In our case $D=3$. To determine the value of $p$ we recall Hajicek's theorem \cite{Hajicek} for ergo-surfaces in four dimensional General Relativity: at the fixed points of the $U(1)$ rotational isometry, ergo-surfaces must either touch the horizon or hit the black hole singularity. Thus, for the double Kerr solution, we expect that at the symmetry axis ($\rho=0$ in Weyl canonical coordinates) the ergo-spheres will only touch when the horizons touch. Hence the merger should be at $\rho>0$ and therefore $p=1$ due to the $U(1)$ symmetry. Thus, the expected angle at the merger point is $\theta_m=\pi/2$ for the double Kerr system, and we shall indeed show this to be the case.

In the co-rotating Kerr solution one can use standard instanton methods to study the thermodynamics of the system since, unlike the counter-rotating case, both the temperatures and the chemical potentials (angular velocities) of the two black holes match.
For the single Kerr solution it is well known  \cite{Davies} that there is a second order phase transition at $J_c/M^2=0.68$ or, in terms of the temperature and angular velocity, at $\Omega_c/T=5.85$. At this phase transition, the specific heat at constant $J$ has an infinite discontinuity, being negative for slowly rotating black holes ($\Omega<\Omega_c$) and positive for fast spinning ones 
($\Omega>\Omega_c$). We shall show that in the double Kerr system there is a similar behaviour. In particular, when considering the canonical ensemble, fixing the total angular momentum, the double
Kerr system is thermodynamically stable for fast spinning black holes. The system is then described by its temperature, total angular momentum and distance between the black holes. 
We shall see that,  provided the angular momentum is large enough, the black holes will evolve to approximate one another, until a minimum distance is reached. At this critical point, which occurs before the horizons merge, the specific heat diverges and the thermodynamical approximation breaks down. One then expects large entropy, and therefore horizon area,  fluctuations to occur. In this situation it is a very interesting question to understand what happens to the system, even before the horizons merge. 

Another question that can be analysed in the framework of gravitational thermodynamics is that of the maximal energy that can be extracted from the collision of two Kerr black holes, which occurs for adiabatic processes. In the case of co-rotating black holes, we shall confirm the bound of 29\% derived by Hawing \cite{Hawking}, studying the dependence of this energy on the angular momentum and distance between the black holes.

This paper is organised as follows. In section \ref{skerr} we shall describe the rod structure of the Kerr solution and its ergo-sphere in Weyl canonical coordinates, since these will be used for the double-Kerr solution. In particular, we shall develop an intuition about the shape of the ergo-sphere in these coordinates. We shall moreover describe how one can construct embedding diagrams for the horizon. In section \ref{sdkerr} we briefly describe the double Kerr solution and its rod structure. In section \ref{skerrcount} we analyse the counter-rotating case. We shall review the main results of \cite{HR} and introduce a suggestive notation. Moreover, we show that the ergo-spheres of the two individual counter-rotating Kerr black holes never merge and we construct the embeddings of the horizon in 3-dimensional Euclidean space as shown in the diagrams of Fig. \ref{embed}. These embedding diagrams give an intuition about the horizon deformations due to the strut and the interaction between the two black holes. In section \ref{skerrco} we analyse the co-rotating Kerr black holes case in a form which is similar to that of the counter-rotating case. Moreover, we exhibit the angle $\theta_m=\pi/2$ at the ergoregions merger point. We plot numerically the behaviour of the angular velocity of the horizon of both black holes as a function $\zeta$ for fixed $M,J$, for various values of $J$, which exhibits the ``slow down'' effect we have described above. A numerical plot of the extremality condition $J_{ext}/M^2$, as function of the coordinate distance $\zeta$ is also given, again exhibiting the behaviour discussed: for fixed $M$,  $J_{ext}/M^2$ increases from unity, as we decrease $\zeta$, to two. In section \ref{horgeo} we analyse the thermodynamics of the co-rotating double Kerr system. We start by computing the action of the double Kerr instanton, therefore computing the free energy in the grand-canonical ensemble. We then consider the canonical ensemble and analyse  thermodynamical stability. Finally, we move to the micro-canonical ensemble to study energy extraction in the process of merging two Kerr black holes.


\section{\label{skerr}Kerr Solution}
In this paper we shall analyse four dimensional, asymptotically flat spacetimes in Weyl canonical coordinates $(t,\phi,\rho,z)$. The line element is of the form
\[ds^2=g_{tt}dt^2+2g_{t\phi}dtd\phi+g_{\phi\phi}d\phi^2+e^{2\nu(\rho,z)}(d\rho^2+dz^2) \ , \]
where $g_{ab}=g_{ab}(\rho,z)$. The $\phi$ coordinate has periodicity $2\pi$ and is chosen such that $g_{t\phi}\rightarrow 0$ at infinity. That is, at infinity we have a non-rotating frame. Thus, ergo-surfaces, if they exist, will always be defined by $g_{tt}=0$. 

\begin{figure}
\includegraphics[width=2.5in]{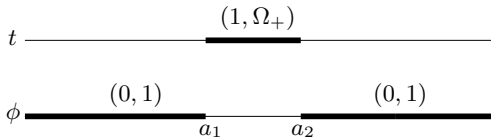}
\begin{picture}(0,0)(0,0)
\put(-108,37){${(1,\Omega_+)}$}
\put(-150,8){${(0,1)}$}
\put(-50,8){${(0,1)}$}
\put(-116,-5){${a_1}$}
\put(-81,-5){${a_2}$}
\put(-188,29){${t}$}
\put(-189,1){${\phi}$}
\end{picture}
\caption{\label{rodkerr} Rod structure for the single Kerr solution.}
\end{figure}

Weyl solutions are conveniently described by their rod structure, including the direction of the rods \cite{Harmark}. The rod structure of the Kerr solution is given in Fig. \ref{rodkerr}. The solution may be obtained by a 2-soliton transformation, at $z=a_1$ (anti-soliton) and $z=a_2$ (soliton) with BZ vectors 
\[\left(1,2a_1 b\right) \ \ {\rm and}\ \  \left(1,2a_2 c\right) \ , \] respectively, from the Schwarzschild solution. The antisymmetric choice $b=-c$ immediately guarantees that the NUT charge is zero and therefore that the solution is asymptotically flat. The result is a two parameter family of vacuum solutions, with parameters given by the length of the temporal rod $a_{21}$ ($a_{ij}\equiv a_i-a_j$) and $b=-c$. 

It turns out that in terms of the \textit{physical} parameters of the solution
\[b=-\frac{a}{r_+}=-c \ , \]
and
\[
a_{21}=r_+-r_-=2\sqrt{M^2-a^2} \  ,\]
 where 
\[r_\pm\equiv M\pm\sqrt{M^2-a^2} \ , \] 
and $M$ and $J=Ma$ denote respectively the ADM mass and angular momentum of the Kerr black hole.

\begin{figure}
\includegraphics[width=3in]{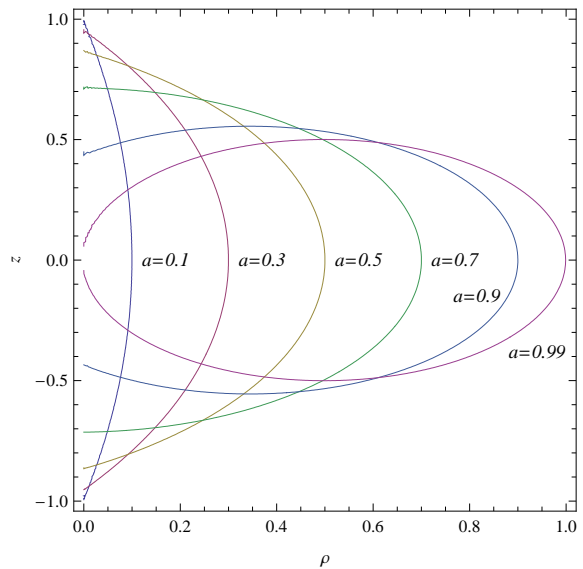}
\begin{picture}(0,0)(0,0)
\end{picture}
\caption{\label{kerrergo} Kerr black hole ergo-surface in Weyl canonical coordinates. We fixed $M=1$ and varied the angular momentum $J=aM$.}
\end{figure}

For future reference it is also useful to recall that the area and temperature of the Kerr black hole are compactly written as:
\[ A=8\pi Mr_+ \ , \qquad T=\frac{r_+-r_-}{8\pi Mr_+} \ . \] 
Note that $AT=a_{21}$. This is a generic result that also holds for individual black holes in the double Kerr solution. A similar statement is true for Smarr's formula
\begin{equation} M=\frac{TA}{2}+2\Omega_+ J \ . \label{smarr} \end{equation}
It follows from Smarr's formula that
the angular velocity of the outer Kerr event horizon is  
\[ \Omega_+=\frac{a}{2Mr_+} \ . \]

\begin{figure}
\includegraphics[width=1.5in]{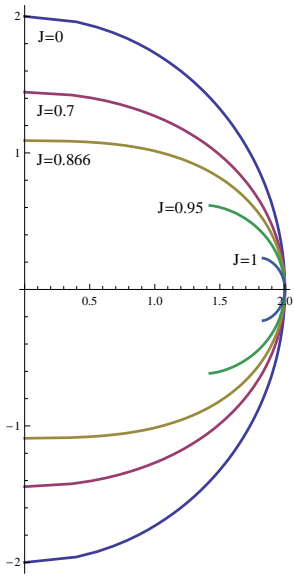}
\begin{picture}(0,0)(0,0)
\end{picture}
\caption{\label{kerrhor} Profile of the Kerr black hole horizon embedding, for fixed mass  $M=1$ and different values of the angular momentum $J=aM$.}
\end{figure}

In Weyl canonical coordinates the Kerr ergo-sphere is defined by the equation
\begin{equation}
\left(\frac{a^2}{r_+^2} - 1\right)^2 (\rho^2 + \mu_1 \mu_2)^2 -
  \frac{4a^2\rho^2}{r_+^2} (\mu_1 - \mu_2)^2 = 0 \ ,
\end{equation}
where
\[\mu_k\equiv \sqrt{\rho^2+(z-a_k)^2}-(z-a_k) \ . \]
In Fig. \ref{kerrergo} we show the location of the Kerr ergo-sphere in Weyl canonical coordinates, for various values of $a$. The location of the (spatial sections) of the outer horizon is, on the other hand, given by $\rho = 0$  and $a_1\le z\le a_2$. Introducing a polar coordinate $\theta$ by
\[ 2z=a_1+a_2+a_{21}\cos\theta \ , \]
the induced metric on the horizon becomes
\begin{equation} ds^2_H=A\left[g(u)d\phi^2+\frac{du^2}{g(u)}\right] \ , \qquad u\equiv \cos\theta \ . \label{horizon}\end{equation}
The conformal factor is constant, $A=2mr_+$, and
\[
g(u)=(1-u^2)\frac{1+b^2}{1+b^2u^2} \ . \]
The above 2-geometry can be embedded in flat Euclidean 3-dimensional space using the embedding functions
\begin{equation} X+iY=\sqrt{Ag(u)}e^{i\phi} \ , \qquad Z=\sqrt{\frac{A}{g(u)}\left(1-\frac{g'(u)^2}{4}\right)} \ , \label{embfunctions} \end{equation}
as long as $g'(u)^2\le 4$ (prime denotes derivative). This always holds in the equatorial plane, for any Kerr black hole; but does not hold for sufficiently fast spinning Kerr black holes and for sufficiently high latitudes \cite{Smarr}. In Fig. \ref{kerrhor} we exhibit the profile of the embedding, for fixed mass and different values of the angular momentum.

\section{\label{sdkerr}Double Kerr solution}
\begin{figure}[t]
\includegraphics[width=2.5in]{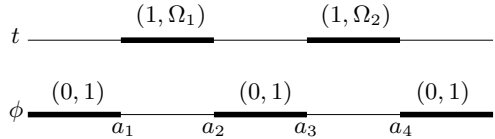}
\begin{picture}(0,0)(0,0)
\put(-141,37){${(1,\Omega_1)}$}
\put(-71,37){${(1,\Omega_2)}$}
\put(-173,8){${(0,1)}$}
\put(-102,8){${(0,1)}$}
\put(-35,8){${(0,1)}$}
\put(-150,-5){${a_1}$}
\put(-116,-5){${a_2}$}
\put(-81,-5){${a_3}$}
\put(-45,-5){${a_4}$}
\put(-188,29){${t}$}
\put(-189,1){${\phi}$}
\end{picture}
\caption{\label{roddoublekerr} Rod structure for the asymptotically flat double Kerr solution, obeying the axis condition, for equal mass black holes with equal ($\Omega_1=\Omega_2$) or opposite ($\Omega_1=-\Omega_2$) spin.}
\end{figure}

The double Kerr solution \cite{KN} is a seven parameter family of solutions of the vacuum Einstein equations. Physically the seven parameters may be written as $M_1,M_2,J_1,J_2,\zeta,M_{axis}$ and $b_{NUT}$, which are respectively the two black hole masses and angular momenta, the coordinate distance between the two black holes, the mass of the axis between them and the NUT charge (see \cite{HR} for a more detailed description). Here we shall focus on asymptotically flat solutions ($b_{NUT}=0$), obeying the axis condition ($M_{axis}=0$) and with equal mass for the black holes $M\equiv M_1=M_2$. Moreover we shall either consider $J\equiv J_1=J_2$, the \textit{co-rotating case} or $J\equiv J_1=-J_2$, the \textit{counter-rotating case}. This leaves us with two 3-parameter families of solutions $(M,J,\zeta)$. For both these families, the rod structure is displayed in Fig. \ref{roddoublekerr}. The coordinate distance is defined as
\begin{equation}
\zeta\equiv a_{32}+\frac{a_{43}+a_{21}}{2} \ . \label{zeta}\end{equation}
We shall refer to the black holes with horizon at $a_1<z<a_2$ and $a_3<z<a_4$ as the first and second black holes, respectively.

\subsection{\label{skerrcount}Counter-rotating Double Kerr}

This case was first considered in \cite{Varzugin} and studied in detail in \cite{HR}. The solution is obtained by a 4-soliton transformation, at $z=a_1$ (anti-soliton), $z=a_2$ (soliton), $z=a_3$ (anti-soliton) and  $z=a_4$ (soliton), respectively with BZ vectors 
\begin{equation} \left(1,2a_1b\right) \ ,  \ \ \left(1,2a_2c\right) \ ,\ \   \left(1,2a_3 c\right)\ \ {\rm and} \ \ \left(1,2a_4 b\right) \ , \label{bzdoublekerr} \end{equation}
 from the double-Schwarzschild solution.  Moreover, we choose from the beginning 
\[a_{21}=a_{43} \ . \]
This symmetric choice of BZ vectors and rod configuration immediately guarantees that the axis condition is obeyed. Additionally, asymptotic flatness is guaranteed if one takes
\begin{equation} a_{21}=a_{43}=\frac{(c+b)(1+bc)}{(c-b)(1-bc)}\,\zeta \ . \label{aflat}\end{equation}
 The result is a three parameter family of vacuum solutions, with parameters given by the BZ parameters $b$ and $c$ and the coordinate distance $\zeta$ (or equivalently $a_{32}$, due to (\ref{zeta})).

\begin{figure}
\includegraphics[width=3in]{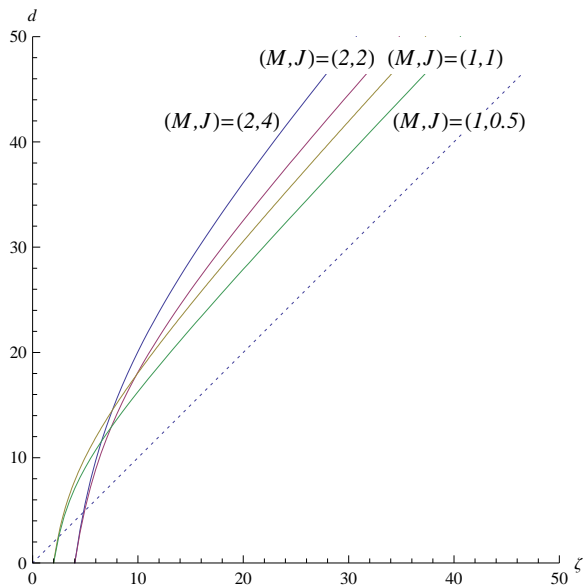}
\begin{picture}(0,0)(0,0)
\end{picture}
\caption{\label{distancecounter} Relation between coordinate distance $\zeta$ and physical distance $d$ in the counter-rotating double Kerr solution, for fixed values of $M,J$. The (dashed) straight line $d=\zeta$ is given for comparison. For $J/M^2<1$, the slope of $d=d(\zeta)$ always approaches unity, asymptotically.}
\end{figure}

For a physical analysis it is desirable to take as independent parameters the Komar mass $M$ of either black hole, the Komar angular momentum $J$ (in modulus) of either black hole and the physical distance $d$ between the black holes. Using the physical distance complicates matters way too much to sustain any hope of finding simple analytical expressions. So, we shall contempt ourselves with taking as independent ``physical'' variables $M$,$J$ and the coordinate distance $\zeta$. This is justifiable since $\zeta$  is a good measure of the physical distance
\[ d(\zeta,M,J)=\int_{a_2}^{a_3} dz\sqrt{g_{zz}}|_{\rho=0} \ , \]
in the sense that, for fixed $M$ and $J$, $d$ grows monotonically with $\zeta$ as shown in Fig. \ref{distancecounter}. Therefore we use the latter, since it simplifies the analysis.

In analogy to the single Kerr solution, we introduce the constants 
\[r_\pm\equiv  M \pm \sqrt{M^2-a^2f(\zeta,M)}\ ,\]
where $M$  denotes the Komar mass of either black hole, $J=Ma$ is (minus) the Komar angular momentum of the first (second) Kerr black hole and 
\[ f(\zeta,M)\equiv \frac{\zeta-2M}{\zeta+2M} \ . \] 
It turns out that in terms of these quantities, the BZ parameters can be simply re-expressed as
\[b=-\frac{af(\zeta,M)}{r_+} \ , \qquad c=\frac{a}{r_+} \ , \]
the constraint (\ref{aflat}) becomes
\begin{equation}
a_{21}=a_{43}=r_+-r_-=2\sqrt{M^2-a^2f(\zeta,M)} \  ,\label{rodsize}\end{equation}
and (\ref{zeta}) takes the form
\begin{equation} a_{32}=\zeta-2\sqrt{M^2-a^2f(\zeta,M)} \ . \label{rodsizedk}\end{equation}

It follows from Smarr's formula and the above results that the angular velocity of the outer Kerr event horizon is given by
\begin{equation} \Omega_1=\frac{a}{2Mr_+}f(\zeta,M)=-\Omega_2 \ , \label{avcounter} \end{equation}
which is relation (\ref{counter1}), given in the introduction. Since the extremal limit is obtained as $a_{21}=a_{43}\rightarrow 0$, it follows from (\ref{rodsize}) that it is given by
\[ \frac{J_{ext}}{M^2}=f(\zeta,M)^{-1/2} \ , \]
which is relation (\ref{counter2}), also given in the introduction.

Finally, the area and temperature of the two black holes can be compactly expressed as 
\begin{equation} A_1=A_2=8\pi Mr_+\left(1+\frac{2M}{\zeta}\right) \ , \label{acounter} \end{equation}
\begin{equation} T_1=T_2=\frac{r_+-r_-}{8\pi Mr_+}\left(1+\frac{2M}{\zeta}\right)^{-1} \ . \label{tcounter}\end{equation}
Notice that the introduction of the constants $r_{\pm}$ displays a remarkable similarity between the physical quantities in this counter-rotating double Kerr solution and the ones of the single Kerr solution exhibited in the previous section.

\begin{figure}
\includegraphics[width=3.1in]{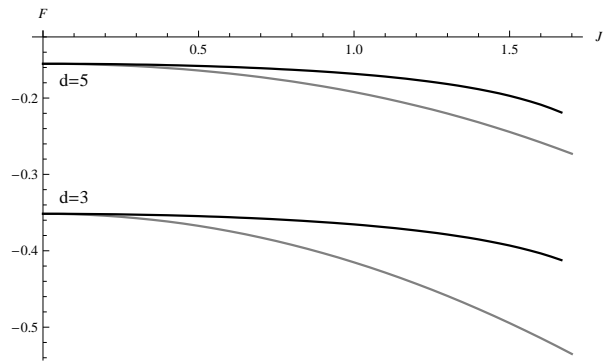}
\begin{picture}(0,0)(0,0)
\end{picture}
\caption{\label{distancej} Force between the two black holes for two physical distances $d=3$ and $d=5$ as a function of $J$ for $M=1$, for the counter-rotating (grey lower curve for each distance) and co-rotating cases (black upper curve for each distance).}
\end{figure}

It follows from (\ref{rodsizedk}) that  $\zeta\rightarrow 2M$ is the horizons \textit{touching limit}, since  $a_{32}\rightarrow 0$. Naively it seems that in this limit the rod structure in Fig. \ref{roddoublekerr} reduces to that in Fig. \ref{rodkerr} with $\Omega_+=0$, by virtue of (\ref{avcounter}).  In other words, that the double Kerr solution with fixed $M$ and $J$ interpolates between two counter-rotating Kerr black holes (at $\zeta\rightarrow \infty$) and a single Schwarzschild black hole (at $\zeta=2M$). Indeed, in the latter limit, the temperature (\ref{tcounter}) yields that of a single Schwarzschild black hole with mass $2M$, and the \textit{total} area of the two black holes $A_1+A_2$ (\ref{acounter}) yields that of a single black hole with mass $2M$. However, in general, this is not the case, because the BZ vectors of the solitons at $z=a_2=a_3$ do not become trivial, cf. (\ref{bzdoublekerr}). Thus, only if one takes $J\rightarrow 0$, the touching limit becomes a merging limit, and one recovers the (single) Schwarzschild solution; otherwise the resulting geometry is singular. 

The force between the two black holes, computed from the massless strut between them, is exactly given by
\begin{equation} F=-\frac{M^2}{\zeta^2}\left[1-\left(\frac{2M}{\zeta}\right)^{2}\right]^{-1} \ .\label{forcecounter} \end{equation}
Naively this is independent of $J$. However, if expressed in terms of the physical distance $d$, rather than the coordinate distance $\zeta$, the force will depend on $J$. One expects that the spin-spin force should be attractive if the black holes are counter-rotating (see \cite{HR} for a discussion of this point). Thus, one expects the modulus of the force to increase as a function of $J$, for fixed physical distance. Although this is a naive expectation (cf. discussion in the next section) it is indeed verified in Fig. \ref{distancej}.

\begin{figure}
\includegraphics[width=3in]{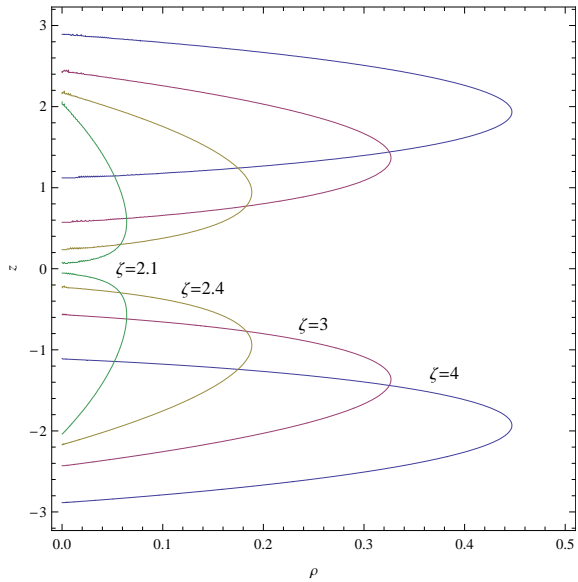}
\begin{picture}(0,0)(0,0)
\end{picture}
\caption{\label{counterfig1} Double Kerr ergo-surface in Weyl canonical coordinates. We fixed $M=1$, $J=0.8$ and varied the coordinate distance as indicated.}
\end{figure}

\begin{figure}
\includegraphics[width=3in]{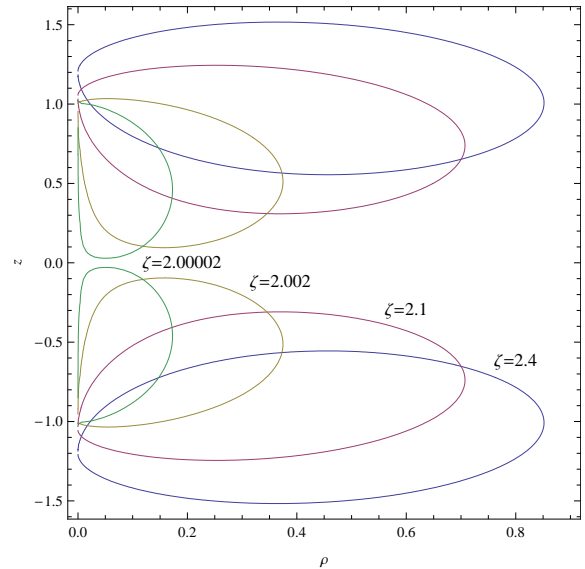}
\begin{picture}(0,0)(0,0)
\end{picture}
\caption{\label{counterfig3} Double Kerr ergo-surface in Weyl canonical coordinates. We  fixed $M=1$ and varied both the angular momentum and the coordinate distance (as indicated) such that as to obey the extremality condition (\ref{counter2}).}
\end{figure}

In Weyl canonical coordinates, the location of the ergo-surface is determined by (taking into account (\ref{aflat}))
\begin{equation}
\{(r_+-r_-)^2\chi +[a(f+1)]^2 \eta\}^2 -[a(f+1) (r_+-r_-) \varepsilon]^2 = 0 \,,
\end{equation}
where $\chi$, $\eta$ and $\varepsilon$ are the following functions of $\rho$ and $z$:
\begin{eqnarray}
 \chi &=& \mu_{13} \mu_{24}\rho^2_{12}\rho^2_{34} \ , \nonumber\\
  \eta &=& \mu_{12}\mu_{34}\rho^2_{13}\rho^2_{24} \ , \\
  \varepsilon &=& \mu_{14}\mu_{23}\rho \left[\mu_{24}\rho^2_{13}-\mu_{13}\rho^2_{24}\right]\ ,\nonumber 
\label{quantities} \end{eqnarray}
and we have used the shorthand notation
\begin{equation}
\mu_{ij}\equiv \mu_i-\mu_j \ , \qquad \rho^2_{ij}\equiv \rho^2+\mu_i\mu_j \ . \label{shorthand}\end{equation}
In Figs. \ref{counterfig1} and \ref{counterfig3} we display the location of the counter-rotating double Kerr ergo-surface. From the plots it is clear that the ergo-spheres of the two individual black holes never touch.

\begin{figure}[h!]
\includegraphics[width=3.0in]{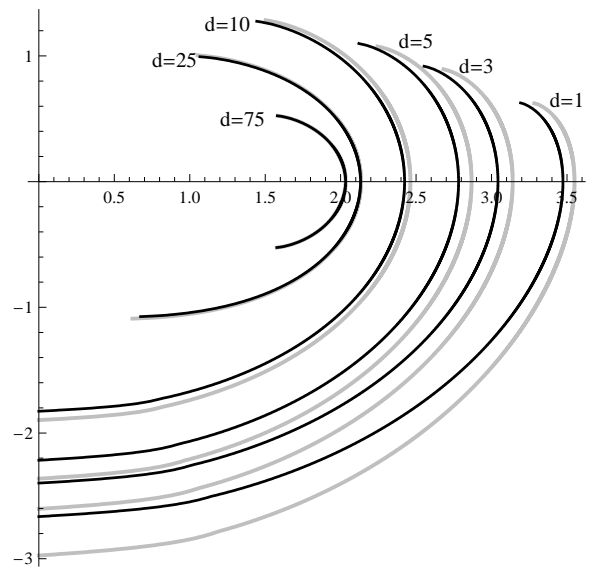}
\begin{picture}(0,0)(0,0)
\end{picture}
\caption{\label{dkerrhor} Profile of the horizon embedding for the first (``lower'') black hole in the counter-rotating (grey lines) and co-rotating (black lines) double Kerr systems. We take $M=J=1$ and various values for the physical distance $d=75,25,10,5,3,1$.}
\end{figure}

As for the single Kerr case one can show that the induced metric on the horizon of each black hole takes the form (\ref{horizon}). One can then use functions (\ref{embfunctions}) to construct an embedding for the horizon. In Fig. \ref{dkerrhor} we display the profile of such embeddings for the first black hole and for various physical distances (fixing $M$ and $J$). One notices that, unlike the single Kerr case, the embedding is not symmetric, as expected from the existence of a strut in one of the sides (upper), as well as the interaction with the other black hole. Since we have chosen, $M=1=J$, these are extremal black holes at infinity. Thus, the embedding covers only a small part of the horizon. However, as they approach, they slow down considerably and the size of the embedding surface increases. In particular, in the ``south pole'', where there is no strut, the horizon becomes completely covered by the embedding. By making a revolution of these profiles we obtain the 3-dimensional embeddings displayed in Fig. \ref{embed}.

\subsection{\label{skerrco}Co-rotating Double Kerr}

Two equal Kerr-Newman black holes were discussed in \cite{Manko:94}, where the solution was generated via transformations of the Ernst potential and the issue of force balance was addressed. Herein we generate it using the inverse scattering construction and focus on different physical issues.

\begin{figure}[t!]
\includegraphics[width=3.4in]{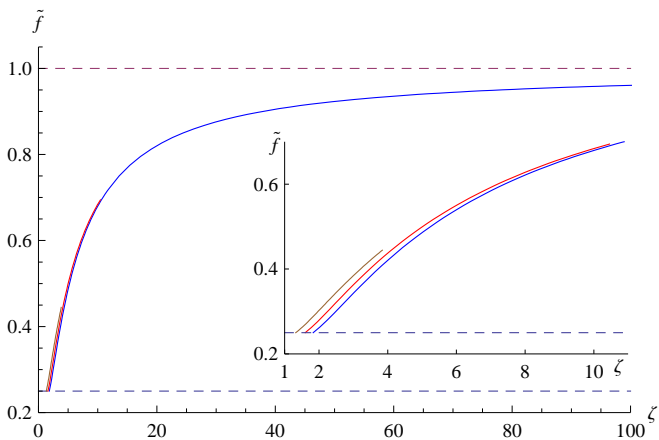}
\begin{picture}(0,0)(0,0)
\end{picture}
\caption{\label{ftilda} $\tilde{f}$ as a function of $\zeta$ for $M=1$ and $J=0.8,1.2,1.5$. The detail shows a zoom near the merging points $\zeta_0$. 
The higher the angular momentum, the lower the maximum allowed value for $\zeta$.}
\end{figure}

The solution is obtained by a 4-soliton transformation, at $z=a_1$ (anti-soliton), $z=a_2$ (soliton), $z=a_3$ (anti-soliton) and  $z=a_4$ (soliton), respectively with BZ vectors 
\begin{equation} \left(1,2a_1 b\right) \ ,  \ \ \left(1,-2a_ 2c\right) \ ,\ \   \left(1,2a_3c\right)\ \ {\rm and} \ \ \left(1,-2a_4b\right) \ , 
\label{bzdoublekerrco} \end{equation}
from the double-Schwarzschild solution.  Moreover, we choose from the beginning 
\[a_{21}=a_{43} \ . \]
This antisymmetric choice of BZ vectors on this symmetric rod structure immediately guarantees that the NUT charge is zero and therefore that the solution is asymptotically flat. Additionally, the axis condition is obeyed provided
\begin{equation} (1+2c^2+b^2c^2)\,\frac{a_{21}^2}{\zeta^2}+2c\,\frac{1-b^2c^2}{b-c}\frac{a_{21}}{\zeta}=(-1+bc)^2 \ . \label{axisco}\end{equation}
 The result is, again, a three parameter family of vacuum solutions, with parameters given by the BZ parameters $b$ and $c$ and the coordinate distance $\zeta$ (or equivalently $a_{32}$, due to (\ref{zeta})). Note that the relation between coordinate and physical distances is qualitatively very similar to the one for the counter-rotating case displayed in Fig. \ref{distancecounter}.

In this case we were not able to express $b$ and $c$ in terms of the ``physical'' parameters $(M,J,\zeta)$. The reason can be traced back to the fact that (\ref{axisco}) gives an expression for $a_{21}=a_{21}(b,c,\zeta)$ less tractable than the one obtained from (\ref{aflat}). Replacing back in the expressions for the Komar quantities one obtains $M=M(b,c,\zeta)$ and $J=J(b,c,\zeta)$, which could be inverted in the counter but not in the co-rotating case.

\begin{figure}[t!]
\includegraphics[width=3.4in]{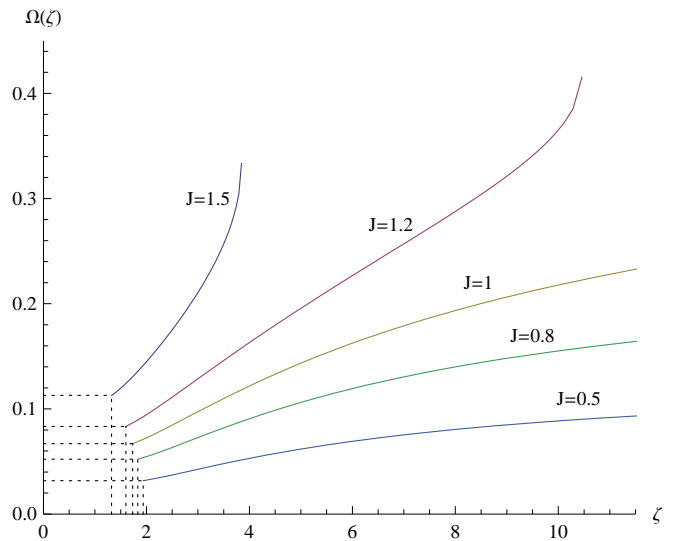}
\begin{picture}(0,0)(0,0)
\end{picture}
\caption{\label{avcofig} Angular velocity, for $M=1$ and fixed $J$ in terms of the coordinate distance in the co-rotating double Kerr solution.}
\end{figure}

Nevertheless, a similar structure to that of the counter-rotating case can be displayed, in terms of an analogue function $\tilde{f}(\zeta,M,J)$, which we determine numerically. That is, defining
\[\tilde{r}_\pm\equiv  M \pm \sqrt{M^2-a^2\tilde{f}(\zeta,M,J)}\ , \] 
where $M$ and  $J=Ma$  denote the Komar mass and Komar angular momentum of either black hole, the physical quantities are expressed by identical expressions to the ones in the last section. Thus, the constraint (\ref{axisco}) becomes
\begin{equation}
a_{21}=a_{43}=\tilde{r}_+-\tilde{r}_-=2\sqrt{M^2-a^2\tilde{f}(\zeta,M,J)} \  ,\label{rodsizeco}\end{equation}
and (\ref{zeta}) takes the form 
\begin{equation} a_{32}=\zeta-2\sqrt{M^2-a^2\tilde{f}(\zeta,M,J)} \ . \label{rodsizedkco}\end{equation}
The merging limit corresponds to
\[
\zeta\rightarrow \zeta_0 \equiv \sqrt{(2M)^2-a^2} \ . \]
Thus, the behaviour of $\tilde{f}(\zeta,M,J)$ in this limit and in the asymptotic limit is
\[
\lim_{\zeta\rightarrow \infty}\tilde{f}(\zeta,M,J)=1 \ , \ \  \lim_{\zeta\rightarrow \zeta_0}\tilde{f}(\zeta,M,J)=\frac{1}{4} \ . \]
The interpolating behaviour is displayed in Fig. \ref{ftilda}.

\begin{figure}
\includegraphics[width=3.4in]{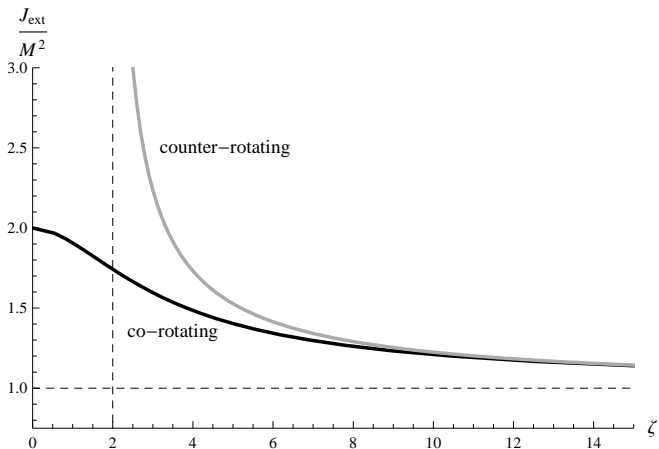}
\begin{picture}(0,0)(0,0)
\end{picture}
\caption{\label{extremal} $J_{ext}/M^2$ in terms of the coordinate distance, for the co-rotating case and the counter-rotating case. We fixed $M=1$ which determines the merging limit in the counter-rotating case to be at $\zeta=2$.}
\end{figure}

It follows from Smarr's formula and the above results that the angular velocity of the outer Kerr event horizon is given by
\begin{equation} \Omega_1=\frac{a}{2M\tilde{r}_+}\tilde{f}(\zeta,M,J)=\Omega_2 \ . \label{avco} \end{equation}
The angular velocity is shown graphically in Fig. \ref{avcofig}. One notes that, for fixed $M$, the angular velocity $\Omega$ decreases with decreasing distance, as anticipated in the introduction. Since the extremal limit is obtained as $a_{21}=a_{43}\rightarrow 0$, it follows from (\ref{rodsizeco}) that it is given by
\[ \frac{J_{ext}}{M^2}=\tilde{f}(\zeta,M)^{-1/2} \ . \]
The comparison between the extremal limit in the counter and co-rotating cases is given in Fig. \ref{extremal}. One observes that for extremal black holes the ratio $J_{ext}/M^2$ increases with decreasing $\zeta$ for both cases, as a consequence of the rotational dragging slow down. However, whereas it diverges in the counter-rotating case, it converges to two in the co-rotating case, corresponding to the $J_{ext}/M^2$ value of the Kerr black hole with mass $2M$ and angular momentum $2J$, that is formed after merging.

\begin{figure}
\includegraphics[width=3.4in]{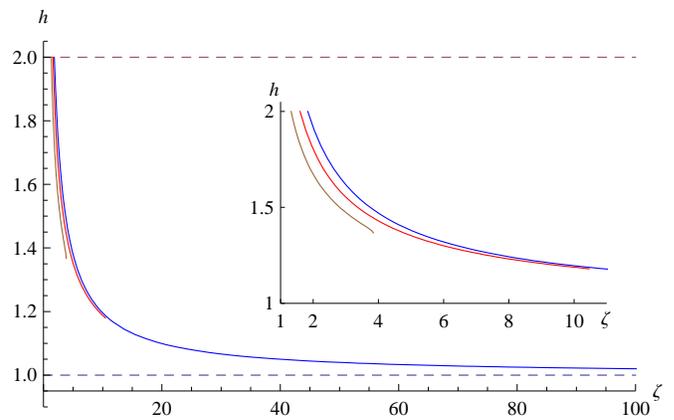}
\begin{picture}(0,0)(0,0)
\end{picture}
\caption{\label{h} $h$ as a function of $\zeta$ for $M=1$ and $J=0.8,1.2,1.5$. The detail shows a zoom near $\zeta=\zeta_{0}$. The higher the angular momentum, the lower the maximum allowed value for $\zeta$.}
\end{figure}

The area and temperature of the two black holes can be compactly expressed as 
\begin{equation} A_1=A_2=8\pi M\tilde{r}_+h(\zeta,M,J) \ , \label{aco} \end{equation}
\begin{equation} T_1=T_2=\frac{\tilde{r}_+-\tilde{r}_-}{8\pi M\tilde{r}_+}h(\zeta,M,J)^{-1} \ . \label{tco}\end{equation}
The function $h$ is exhibited in Fig. \ref{h}.  Note that it has the same asymptotic behaviour, at infinity and in the merging limit, as the corresponding function in the counter-rotating case, cf. (\ref{acounter}) and (\ref{tcounter}).

The force between the two black holes, for fixed mass and physical distance, is displayed in Fig. \ref{distancej}. In this case the naive expectation is that the force should \textit{decrease} in modulus, since by general arguments the co-rotating spin spin force should be repulsive. Fig.  \ref{distancej} shows, however, that this is not the case. The force increases, in modulus, as we increase $J$, for fixed mass and physical distance. 

The breakdown of our naive expectation is, most likely, associated with finite size effects. As we vary $J$, for fixed $d$ and $M$, we are varying the horizon shape and therefore higher gravitational multiple moments should contribute to the force at the same order as the spin-spin interaction. The plot in Fig. \ref{distancej} is informing us that they mask the spin-spin force completely.

It is, however, reassuring that the force is greater, in modulus, in the counter-rotating case than in the co-rotating one. Indeed, it has been argued by Hawking \cite{Hawking} that more energy can be extracted in a head on collision of Kerr black holes, if the spins are anti-aligned, rather than aligned, with the direction of separation. Taking one of the black holes to be much smaller than the other, Hawking derived a spin-spin interaction energy which was later shown by Wald \cite{Wald}, using a test particle (with spin) approximation, to be the work of the spin-spin force between the black hole and the spinning test particle. Such force is attractive (repulsive) for anti-aligned (aligned) spins. 

The fact that our analysis unveils a force which is more attractive in the counter-rotating than the co-rotating case is in qualitative agreement with the Hawking-Wald spin-spin interaction energy. But note that, in any case, there is no overlap between their analysis and ours. We consider two equal mass and equal spin (in modulus) black holes whereas their calculation is limited to a test object in the background of a Kerr black hole. This  means that the finite size effects we are including are completely neglected in their approximation.

\begin{figure}
\includegraphics[width=3.5in]{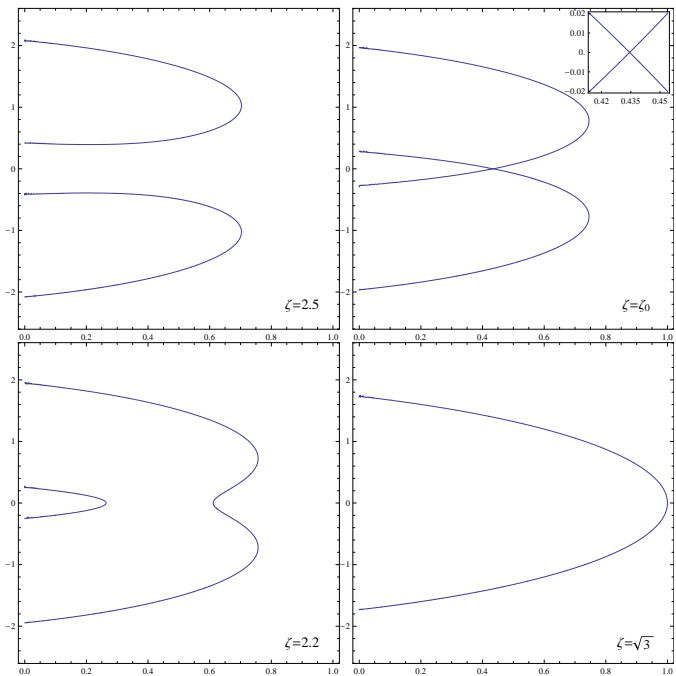}
\begin{picture}(0,0)(0,0)
\end{picture}
\caption{\label{cofig1} Evolution of the co-rotating double Kerr ergo-surface in Weyl canonical coordinates. We fixed $M=1=J$, and considered various distances. The ergosurface merging happens for $\zeta=\zeta_0=2.243464$, while the horizon merging for $\zeta=\sqrt{3}$. The detail in the second figure shows a zoom of the merger, in a box of equal length in $\rho$ and $z$ so that the $\pi /2$ angle becomes clear.}
\end{figure}

In Weyl canonical coordinates, the ergo-surfaces for the co-rotating double Kerr are defined, in terms of $b,c$ and $\zeta$, by the following equation (taking into account (\ref{axisco})):
\[
\begin{array}{c}
\left\{\left[(1+bc)^2\chi-(b-c)^2\eta\right]\rho^2_{23}\rho^2_{14}-4bc\alpha\right\}^2 \\ \\ -[b(1-c^2)y-c(1-b^2)x]^2=0 \ , \end{array} \]
where we are using the definitions (\ref{quantities}), the shorthand notation (\ref{shorthand}) and
\begin{eqnarray*}
 \alpha &=& \mu_{13} \mu_{24}\rho^2_{13}\rho^2_{24}\left[\mu_{23}\mu_{14}\rho^2+\rho^2_{23}\rho^2_{14}\right]\ , \nonumber\\
  x &=& \rho\mu_{23}\rho^2_{14}\left[\mu_{14}^2\rho^2_{22}\rho^2_{33}-\chi-\eta\right] \ , \nonumber \\
  y &=& \rho\mu_{14}\rho^2_{23}\left[\mu_{23}^2\rho^2_{11}\rho^2_{44}-\chi-\eta\right] \ .
\label{newquantities} \end{eqnarray*}
In Fig. \ref{cofig1} we display the merging of the ergo-surfaces for (asymptotically) extremal black holes and exhibit the merger angle of $\pi/2$.

As for the counter rotating double Kerr one can show that the induced metric on the horizon of each black hole takes the form (\ref{horizon}). One can then use functions (\ref{embfunctions}) to construct an embedding for the horizon. In Fig. \ref{dkerrhor} we display the profile of such embeddings for the first black hole. They are quite similar to the counter-rotating case, but the embedding covers a slightly larger portion of the horizon, for small distances. The 3-dimensional embeddings are given in Fig. \ref{embed}.


\section{\label{horgeo}Thermodynamics for the co-rotating Double Kerr}

For the counter rotating double Kerr the two black holes have the same temperature but not the same chemical potential (angular velocity). 
Thus, equilibrium thermodynamics can only be meaningfully considered in the co-rotating case. Let us therefore study the co-rotating 
double Kerr black hole as a thermodynamical system. 

\subsection{\label{grand}Grand canonical ensemble}

First we consider the grand canonical ensemble with temperature $T$ and chemical potential given by the angular velocity $\Omega$. 
The free energy can be computed from the standard Euclidean quantum gravity approach. One considers geometries satisfying
the periodic boundary condition $(\tau,z,\rho,\phi)\equiv (\tau+\beta,z,\rho,\phi+i\beta\Omega)$, where $\tau$ is the Euclidean
time and $\beta$ is the inverse of the temperature, and then computes the action of the double Kerr instanton. 
The free energy $W=W(T,\Omega)$ is then related to the action by $I=\beta W$.

To find the double Kerr  instanton, first one imposes the existence of two horizons separated
by a coordinate length $L$ along the rotational symmetry axis of the Weyl canonical coordinates, for instance by choosing the rod structure
introduced in section \ref{sdkerr}.
Then, after solving Einstein's equations, one
finds that the geometry has a strut singularity between both black holes.
Thus, the thermodynamical analysis is performed at fixed coordinate length $L$ of the strut between both Kerr black holes.
This variable is related to $\zeta$ and the parameters in the rod structure of section \ref{sdkerr} by
$L = a_{32}=\zeta - a_{43}$. It should be remarked that $L$ is a {\em mechanical} parameter, not a thermodynamical one. Indeed, 
a simple computation shows that the entropy density associated to a single strut (or to a cosmic string) is zero. The role of this strut is simply to 
impose the mechanical equilibrium between both holes (one should not think of thermal fluctuations of $L$, as it is the case with  $T$ and $\Omega$). 
With this word of caution we shall write bellow $W=W(T,\Omega,L)$, keeping in mind that $T$ and $\Omega$ are thermodynamical variables, while
$L$ is a mechanical parameter.

The contribution to the action from both black holes arises, as usual, from the boundary term of the Euclidean action at infinity and is given by
$$
\beta\,\frac{M}{2}+\beta\,\frac{M}{2}\,.
$$
There is, however, another contribution to 
the action coming from the conical singularity between the black holes, which were kept at fixed coordinate distance $L$, given by
$$
-\frac{1}{16\pi}\,\int d^4x\sqrt{g}\,R =  - \frac{1}{8\pi}\,A\,\delta\,,
$$
where $\delta=2\pi(1-e^{-\gamma})$ is the deficit angle and $A=\beta \,e^{\gamma}L$ is the area of the surface spanned by this singularity, which is located at
$\rho=0$, $|z|<L/2$ and $0\le \tau < \beta$. The function $\gamma$ appears in the definition of the metric given in \cite{HR} and is computed
at the strut  \footnote{$\gamma=\gamma_{III}$ in the notation of \cite{HR}.}.
Notice that this contribution to the action is positive, since the deficit angle is negative, as the strut exerts 
an ``outwards'' force, therefore preventing the black holes from falling into each other.
Putting everything together we have 
$$
W(T,\Omega,L)=M + \frac{1}{4}\,(1-e^\gamma)L\,.
$$
In this expression one should regard both the mass $M$ and the function $\gamma$ as functions of $T$, $\Omega$ and $L$.  For instance, in
the simpler case of the double Schwarzschild black hole the dependence of the mass on the temperature 
and coordinate distance is given by the relation \cite{CP}
$$
M=\frac{1}{8\pi T}\, \frac{L + 2M}{L + 4M}\,.
$$
In a reversible thermodynamical process the variation of the free energy $W$ is given by
$$
dW = -SdT-\mathcal{J}d\Omega\ ,
$$
where $S$ is the total entropy  and $\mathcal{J}$ is the total angular momentum of the system (\textit{not} the angular momentum $J$
of each black hole, as defined in the previous sections). As we shall see bellow
both quantities are not simply the sum of the entropy and angular momentum of each black hole, since the interaction 
between both holes must be taken into account. 


\begin{figure}
\includegraphics[width=3.4in]{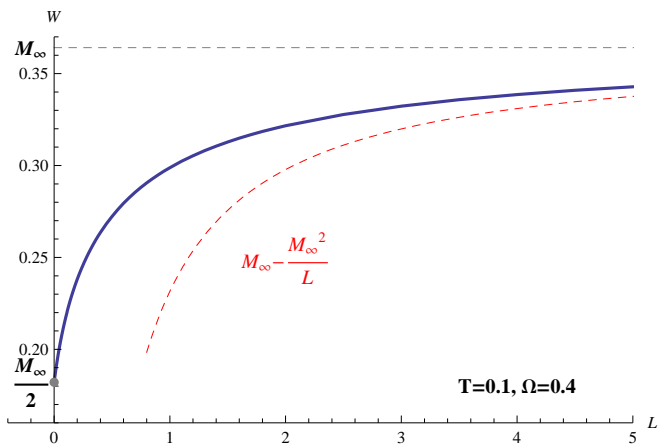}
\begin{picture}(0,0)(0,0)
\end{picture}
\caption{\label{W} Free energy $W$ as a function of $L$ for fixed particular values of $T$ and $\Omega$. 
Asymptotically the free energy matches the  Newtonian gravitational potential energy, while for short distances it is finite.}
\end{figure}

Let us now consider a mechanical process where the black holes start from an infinite distance and then slowly reduce their distance,
while their temperature $T$ and chemical potential $\Omega$ are kept fixed. The mass of each black hole $M_\infty$, when they are
infinitely far apart,  is given by the single Kerr expression
$$
4 \Omega^2 M_\infty^2+8\pi TM_\infty  = 1\,.
$$ 
The asymptotic value of the free energy $W_\infty=M_\infty$ is just the free energy of two isolated Kerr black holes. 
The angular momentum of each black hole at infinite separation $J_\infty$ is also given by the single Kerr value.
As usual, the force acting on the black holes can be computed from the variation of the free energy $W=W(T,\Omega,L)$ with respect to $L$
$$
\mathcal{F} = - \left.\frac{\partial W}{\partial L}\right|_{T,\Omega}\,.
$$
As the black holes slowly move in,
the variation of the free energy  is given by the  work done {\em on}  the system
$$
W- W_{\infty} = - \int_\infty^L \mathcal{F} dL'\,.
$$
In Fig. \ref{W} the generic form of the free energy as  a function of $L$ is shown. The attractive force between the black
holes is doing work during the process, so that the system reduces its free energy $W$. This free energy is nothing but  the
static interaction potential between the black holes, including the spin-spin interaction and the effect of the 
finite size of the black hole horizons (i.e the multipolar force terms).  Notice, however, that during this process the system is exchanging
heat with the thermal bath (in fact losing heat since the entropy decreases), as well as angular momentum, so that the mechanical force here
computed {\em is not} the same mechanical force considered in sections \ref{skerrcount} and \ref{skerrco}. For large distances 
$$
W \simeq M_\infty - \frac{M^2_\infty}{L}\,,
$$
correctly reproducing the Newtonian interaction between two point-like masses. 
As expected, the merging of the black holes is thermodynamically favoured. The final state is then that of a single 
black hole of mass $M_\infty$ and angular momentum $J_\infty$, and the free energy is reduced to $W=M_\infty/2$. 

It is interesting to consider the particular case of zero temperature, which corresponds to the merging
of two extremal black holes with fixed angular velocity $\Omega$. In this case we start with two black holes, each with mass and angular momentum
given by $M_\infty^2 = J_\infty = 1/(4\Omega^2)$. The final state is that of a single black hole with the same mass and angular momentum. 
Thus, just before merging,  each black hole will have mass $M_\infty/2$ and angular momentum $J_\infty/2$. These black holes are extremal and
satisfy the $L\rightarrow 0$ limiting relation $(J_\infty/2)/(M_\infty/2)^2=2$. Thus, from the thermodynamical evolution at  $T=0$ and fixed $\Omega$ we recover the 
result discussed in the introduction and shown in section \ref{skerrco}; in fact this is a special case of the result therein. Indeed, as we approach two black holes 
varying their Komar mass and angular momentum as to keep them extremal at all distances, we may, although we do not have to, keep their angular velocity constant.

\begin{figure}
\begin{center}
\includegraphics[totalheight=2.7in,viewport=0 78 350 417,clip]{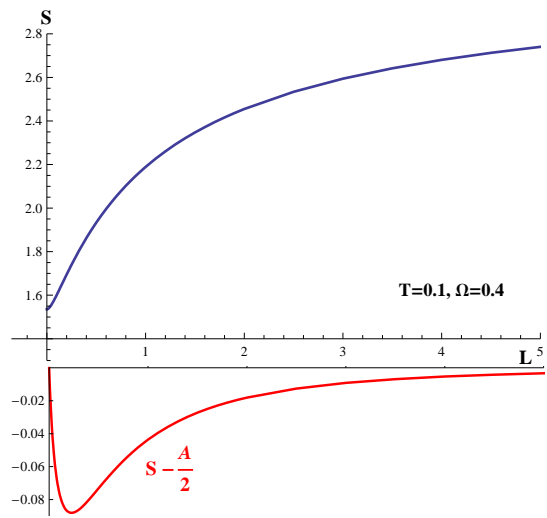}
\end{center}
\vspace{-0.2cm}
\caption{The entropy as a function of  $L$ for fixed particular values of $T$ and $\Omega$.
Shown is also the difference between the entropy and the contribution from each black hole given by the
area law.}
\label{Entropia}
\end{figure}

The total entropy and angular momentum of the system can also be computed from the thermodynamical potential at fixed $L$,
$$
S = - \left.\frac{\partial W}{\partial T}\right|_{\Omega}\,,\ \ \ \ \ \ \ \ 
\mathcal{J} = -  \left.\frac{\partial W}{\partial \Omega}\right|_{T}\,.
$$
In particular, notice that the entropy computed from this expression {\em is not given} by the sum of the entropy of each black hole obtained from the area
law. It contains 
additional terms, which are subleading at large distances,  that arise from the interaction between the black holes, since the interaction term in $W$ also depends
on the temperature. When the black holes merge we recover again the area law. In Fig. \ref{Entropia} we show the entropy of the system as a function of the distance $L$, as well
as the difference between the entropy and the contribution from each black hole obtained from the area
law.  A similar behaviour is observed for the 
total angular momentum $\mathcal{J}$ of the system, so that only when the black holes are infinitely far apart, or when they merge, 
$\mathcal{J}$ is the angular momentum computed from the Komar integral.

\subsection{Review of single Kerr thermodynamical stability}

\begin{figure}
\begin{center}
\includegraphics[height= 1.7in]{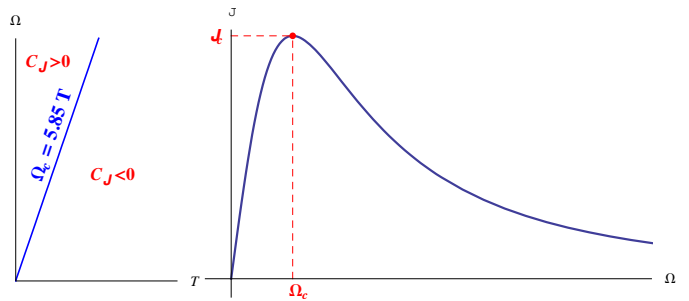}
\end{center}
\caption{Phase diagram (left) and isotherm $J=J(\Omega)$ (right)  for the single Kerr black hole.
The critical angular velocity $\Omega_c/T=5.85$ corresponds to a critical angular momentum obeying $J_c/M=0.68$.
At this critical line the specific heat $C_J$ has an infinite discontinuity, while $\epsilon_T$ changes sign.}
\label{Skerr}
\end{figure}

To analyse the thermodynamical stability of the double Kerr system, let us first recall some results regarding the single Kerr black hole \cite{Davies}. In the grand canonical ensemble, the stability condition is the positivity of the \textit{Weinhold metric} \cite{Weinhold}; its inverse is
\[
g_W^{ab}=-\frac{\partial^2 W}{\partial x^a \partial x^b}\equiv
\left[
\begin{array}{cc}
\beta C_\Omega & \eta\\
\eta & \epsilon_T
\end{array}
\right]\ ,
\]
where $x^a=(T,\Omega)$ (see \cite{JR} for a clear and analogous discussion of the 
thermodynamical stability of the Reissner-Nordstr\"om black hole). The eigenvalues of this matrix are given by
$\beta C_J=\beta C_\Omega-\eta^2/\epsilon_T$ and $\epsilon_T$, where
\[
C_{J} = T\,\left.\frac{\partial S}{\partial T}\right|_{J}\,\ \ \ \ \ 
\epsilon_T =\left. \frac{\partial J}{\partial \Omega}\right|_{T}
\]
are, respectively, the specific heat at constant $J$ (which, for a single black hole, equals $\mathcal{J}$) and the isothermal `moment of inertia'.  Fig. \ref{Skerr} illustrates the behavior of both $C_J$ and $\epsilon_T$ for the single Kerr black hole: there is a 
critical value of the angular velocity 
$$
\frac{\Omega_c}{T} = 5.85 \,,
$$
below which $C_J<0$ and above which  $\epsilon_T<0$, showing that  in the grand canonical ensemble the Kerr black hole
is \textit{always} unstable.

Working instead in the canonical ensemble, where the angular momentum $J$ is fixed, the Kerr black hole has a stable phase.
In this case one only needs to consider the sign of the specific heat $C_J$, concluding that slowly rotating Kerr black holes with $\Omega<\Omega_c$
are thermodynamically unstable, while \textit{fast spinning Kerr black holes with}  $\Omega>\Omega_c$ \textit{are stable}. 
Both phases, schematically represented in Fig. \ref{Skerr}, are divided by a second order phase 
transition, since at the critical value of the horizon angular velocity the free energy has a discontinuous second derivative. 
This critical point is characterised by an infinite discontinuity of the 
specific heat $C_{J}$.
The distinction between the two phases can be clearly seen from the isotherm
$J=J(\Omega)$, also shown in Fig. \ref{Skerr}. Indeed,   the 
divergence in the specific heat occurs when the derivative of the isotherm $J=J(\Omega)$ 
vanishes, which in turn implies that the difference between the specific heats at fixed $\Omega$ and $J$ diverges. 
Since $C_\Omega$ is finite at the transition, $C_J$ will diverge.

\subsection{Canonical ensemble}

\begin{figure}[t]
\begin{center}
\includegraphics[height= 2.2in]{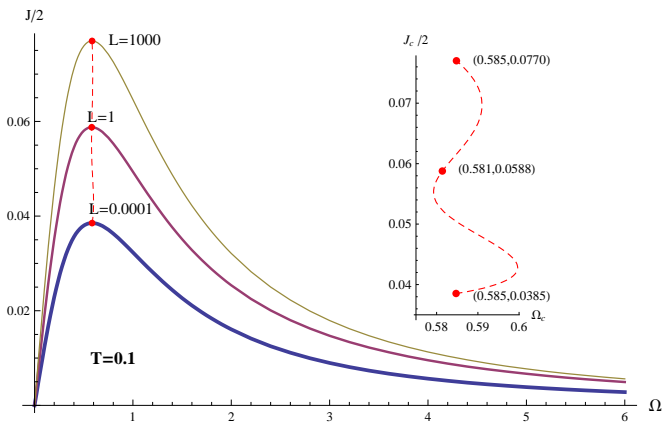}
\end{center}
\caption{Isotherms $\mathcal{J}=\mathcal{J}(\Omega,L)$ for fixed temperature $T$ and three different values of $L$. The detail on the right shows the line of maxima 
of the isotherms with respect to variations of $\Omega$. This line defines the critical angular velocity $\Omega_c=\Omega_c(T,L)$.}
\label{JOmega}
\end{figure}

\begin{figure}[t]
\begin{center}
\includegraphics[height= 2.2in]{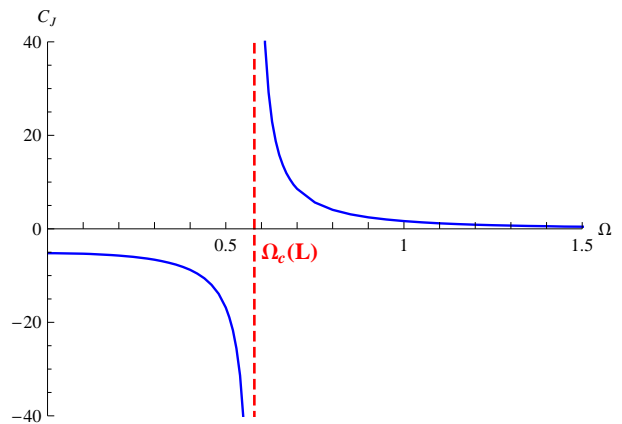}
\end{center}
\caption{Behaviour of the specific heat $C_{\mathcal{J}}$, for a generic value of $L$, as a function of $\Omega$. In this plot we set $T=0.1$ and $L=1$, so that the
discontinuity corresponds to the maxima of the $L=1$ curve of Fig. \ref{JOmega}.}
\label{CJOmega}
\end{figure}

Clearly, just as the single Kerr black hole, the co-rotating double Kerr system is thermodynamically unstable in the grand canonical ensemble. 
Let us, therefore, consider the thermodynamical stability in the canonical ensemble, where the angular momentum $\mathcal{J}$ is held fixed. 
A good starting point is to analyse the behaviour of the isotherms 
$\mathcal{J}=\mathcal{J}(\Omega,L)$, shown in Fig. \ref{JOmega}.
We saw in section \ref{grand} that when two black holes start infinitely far apart, each with angular momentum $J_\infty=\mathcal{J}(\Omega,\infty)/2$, and then move in
at fixed  $T$ and $\Omega$, the final state is that of a single black hole with angular momentum $J_\infty = \mathcal{J}_\infty(\Omega,0)$. 
It follows that $\mathcal{J}(\Omega,\infty)=2\mathcal{J}(\Omega,0)$, as can be verified in Fig. \ref{JOmega}.
Like the single Kerr black hole, the double Kerr system also has two distinct phases, one of which is stable. 
To see this, notice that for any fixed distance $L$ the isotherm $\mathcal{J}=\mathcal{J}(\Omega,L)$ has
a maximum $\mathcal{J}_c$ at some $\Omega_c$. Hence, fixing the temperature $T$, as $L$  varies from $0$ to $\infty$, there is a critical line given by
$$
\frac{\Omega_c}{T} = \omega(T,L)\,,
$$
with the asymptotic behaviour $\omega(T,\infty)=\omega(T,0)=5.85$ given by the single Kerr result. As a function of  $L$ we observe an oscillatory 
behaviour in $\Omega_c$, as shown in the detail of Fig. \ref{JOmega}.
Changing $T$ does not change this qualitative behaviour. 
We conclude that the specific heat $C_\mathcal{J}$ has an infinite discontinuity at the critical line,
which separates the stable, $C_\mathcal{J}>0$, and unstable, $C_\mathcal{J}<0$,  phases. 
As for the single Kerr, both phases are divided by a second order phase 
transition. In Fig. \ref{CJOmega} we show the behaviour of the specific heat $C_\mathcal{J}$ computed for the double Kerr system for a finite distance $L$, 
which is similar to that of the single Kerr.

\begin{figure}[t]
\begin{center}
\includegraphics[height= 2in]{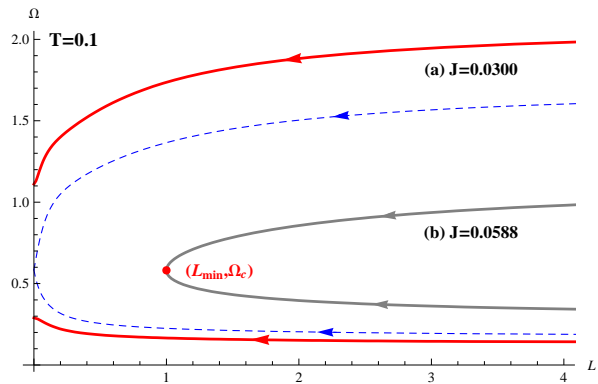}
\end{center}
\caption{The fixed $T$ and $\mathcal{J}$ trajectories in the $\Omega - L$ plane.  The arrows indicate the mechanical evolution of the system given
by the thermodynamical potential $W'$. The red trajectories (a) describe  the merging of  fast rotating (top) and slow rotating (bottom)
Kerr black holes. For the grey  trajectories (b) both fast and slow rotating Kerr black holes evolve to a minimum distance $L_{min}$. Recall that 
the fast rotating black holes are the stable ones. The dashed trajectories correspond to the  limiting case $\mathcal{J}=\mathcal{J}_c(0,T)$.}
\label{PerfilOmegaZ}
\end{figure}

\begin{figure}
\begin{center}
\includegraphics[height= 1.8in]{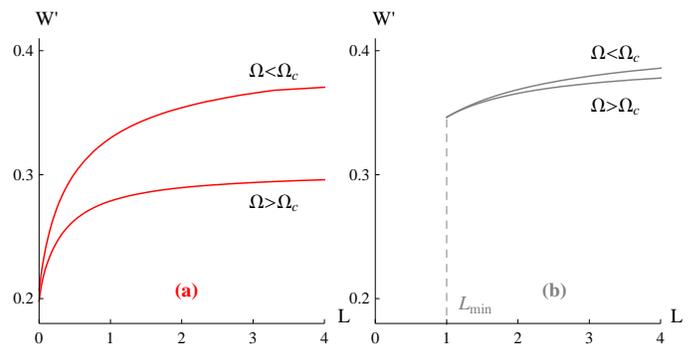}
\end{center}
\caption{Thermodynamical potential $W'$ along the trajectories (a) and  (b) of Fig. \ref{PerfilOmegaZ}.}
\label{WlinhaL}
\end{figure}

For a given value of the angular momentum $\mathcal{J}<\mathcal{J}_c$
the single Kerr black hole system has two possible states, one of which
is stable (see Fig. \ref{Skerr}). For the double Kerr system this structure is richer since there is also the 
distance parameter $L$. From Fig. \ref{JOmega} it is clear that there are  two qualitatively different
cases determined by the value of the angular momentum  $\mathcal{J}$ with respect to its critical value $\mathcal{J}_c=\mathcal{J}_c(L,T)$ at $L=0$.
When the angular momentum of the system $\mathcal{J}$ satisfies  $\mathcal{J} <  \mathcal{J}_c (0,T)$, the coordinate distance $L$ may take values
from $\infty$ to $0$. On the other hand, when $\mathcal{J} > \mathcal{J}_c (0,T)$, it is clear that the black holes may only reach a minimum distance $L_{min}$.
In both cases there are stable and unstable phases. These facts can be seen from  Fig. \ref{PerfilOmegaZ}, where the slices 
$\mathcal{J}=const.$ of the isotherms of Fig. \ref{JOmega}
are represented in the  $\Omega-L$ plane. In particular, given $\mathcal{J}$ such $\mathcal{J}>\mathcal{J}_c(0,T)$,  the black holes will move to the critical line at a finite
distance  $L_{min}$ defined by
$$
\mathcal{J}=\mathcal{J}_c(L_{min},T)\,.
$$
To better understand this statement, let us consider the
thermodynamical potential associated to the mechanical evolution of the system at fixed $T$ and $\mathcal{J}$, which is given by
$$
W'=W+\Omega \mathcal{J}\,.
$$
The black holes will evolve mechanically to minimize  the potential $W'=W'(T,\mathcal{J},L)$  with respect to $L$. This evolution is represented by the arrows in Fig. \ref{PerfilOmegaZ}.
In Fig. \ref{WlinhaL} we show the behaviour of the potential $W'$ as a function
of the distance $L$  for the trajectories 
$(a)$ and $(b)$ of Fig. \ref{PerfilOmegaZ}. Clearly, the force
$$
\mathcal{F} = - \left.\frac{\partial W'}{\partial L}\right|_{T,\mathcal{J}}\,.
$$
is always attractive.

We have seen that for {\em fixed temperature and fixed and high enough angular momentum  Kerr black holes have a minimum allowed distance}! In the particular case of the stable fast rotating
black holes, one reasonable question to ask is: what happens to the system as the minimal distance $L_{min}$ is reached? At this point the
specific heat $C_\mathcal{J} \rightarrow +\infty$, so that the response of the system to thermal fluctuations is very large 
and one can \textit{not} trust the thermodynamical approximation. Fluctuations of the entropy, and therefore of the horizons area, will be very large. 
For this reason it is tempting to conjecture that at this finite distance a phase transition will occur under which a single horizon is formed.

\subsection{Micro-canonical ensemble and energy extraction}

\begin{figure}
\includegraphics[width=3.4in]{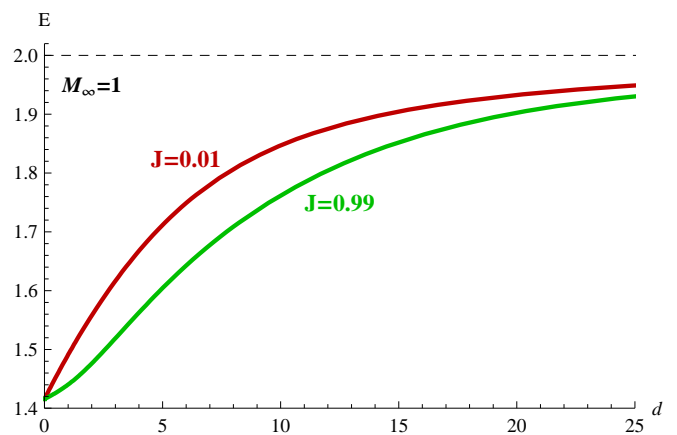}
\begin{picture}(0,0)(0,0)
\end{picture}
\caption{\label{Energy}Energy as a function of the physical distance $d$ for fixed entropy and angular momentum. For both curves the asymptotic value of the
black holes masses is the same.}
\end{figure}

Finally one may also wish to consider the evolution of the double Kerr system at fixed entropy $S$, angular momentum $\mathcal{J}$ and distance $L$. 
The corresponding thermodynamical potential is the total energy of the system $E$ given by
$$
E = W' + TS\,,
$$
with differential
$$
dE= TdS + \Omega d\mathcal{J}\,.
$$
This evolution is relevant to understand the maximal energy that can be extracted from the merging of both black holes for fixed $\mathcal{J}$, since in that case
the variation of the energy is maximised by adiabatic processes. In Fig. \ref{Energy} we show the energy of the double Kerr system for fixed
entropy $S$ and angular momentum $\mathcal{J}$, as a function of the physical distance $d$. Each curve represents a double Kerr system with the same
total energy (asymptotic masses), but different values of the entropy and angular momentum. In both cases we obtain the upper bound of $29\%$ for energy extraction
derived by Hawking \cite{Hawking}, which is independent of the angular momentum. Fig. \ref{Energy} also shows that, for fixed initial energy and until a finite
distance is reached, one may extract more energy from extremal black holes. This fact agrees with the mechanical analysis of the force between 
co-rotating Kerr black holes given in section \ref{sdkerr} and illustrated by Fig. \ref{distancej}, since the higher $J$ the larger is the force in
modulus for fixed mass black holes.

\section{\label{fr}Final Remarks}
In this paper we have studied physical aspects of two special cases of the double Kerr solution, which we have dubbed the counter-rotating and the co-rotating cases. These cases are asymptotically flat, obey the axis condition and have $M_1=M_2$ and $|J_1|=|J_2|$. They simplify dramatically the general double Kerr solution, albeit retaining still a considerable degree of complexity. This is why we have chosen not to present their explicit construction herein. The interested reader is referred to \cite{HR} for details.

An application of this system is to study rotational dragging effects at the level of an exact solution of four dimensional General Relativity, as described at length in the introduction. A striking physical consequence is the possibility that the dimensionless extremality ratio $J_{ext}/M^2$ may exceed unity, already discussed in \cite{HR} for the counter-rotating case. A somewhat analogous phenomenon takes place for a double Reissner-Nordstr\"om solution of two black holes with \textit{opposite} charges \cite{Chandra,Emparan}, for which the extremality ratio $|Q_{ext}| /M$ may exceed unity \cite{EmparanTeo}. A tempting interpretation is that (as for the double Kerr solution) it is the electric potential on the horizon (angular velocity) rather than the total charge (angular momentum) that determines the existence of a horizon. It would be interesting to pursue the comparison between the charged and rotating di-holes in more detail. A clear difference is that charged diholes with the same charge (see e.g. \cite{jap}) yield, in the extremal limit, a Majumdar Papapetrou solution, for which the ratio between charge (in modulus) and mass is one.

Another possible application of these double Kerr systems is to study geodesic motion in the presence of two black holes. This can certainly be done on the equatorial plane of the co-rotating case which (unlike the counter-rotating case) provides a totally geodesic submanifold. A complete study of orbits, however, would require the geodesic problem to be Liouville integrable. This might sound unlikely. But it was recently pointed out by Will \cite{Will} that the Newtonian version of the double Schwarzschild system does posses a hidden symmetry. It would therefore be of interest to see if an irreducible Killing tensor exists in both the double Schwarzschild solution and the particular cases of the double Kerr we have considered.

\section*{Acknowledgments}
We would like to thank R. Monteiro and J. Santos for useful discussions and comments on a draft of this paper. C.H. is funded by a Ciencia 2007 research contract. C.R. is funded by FCT through grant SFRH/BD/18502/2004. CFP is partially funded by FCT through the POCI programme.


\bibliography{prd}

\end{document}